\documentclass[prl,aps,10pt,superscriptaddress,twocolumn,floatfix, nofootinbib]{revtex4-1}
\usepackage{graphicx,color}
\usepackage{amsmath,amssymb,bm}
\usepackage{braket}	

\DeclareMathOperator{\Tr}{Tr}
\usepackage[plainpages=false,pdfpagelabels,colorlinks=true,linkcolor=red,urlcolor=blue,citecolor=blue,pdftitle={Titl}e,pdfauthor={},pdfdisplaydoctitle=true,pdfduplex=DuplexFlipLongEdge]{hyperref}

\usepackage{multibib}
\newcites{SI}{Supplemental Information References}

\begin{document}
\title{Harnessing quantum chaos in spin-boson models for all-purpose quantum-enhanced sensing}
\author{Y. Zhang}
\author{J. Zuniga Castro}
\author{R.~J. Lewis-Swan}
\affiliation{Homer L. Dodge Department of Physics and Astronomy, The University of Oklahoma, Norman, OK 73019, USA }
\affiliation{Center for Quantum Research and Technology, The University of Oklahoma, Norman, OK 73019, USA}
\date{\today}

\begin{abstract}    
Many-body quantum chaos has immense potential as a tool to accelerate the preparation of entangled states and overcome challenges due to decoherence and technical noise. Here, we study how chaos in the paradigmatic Dicke model, which describes the uniform coupling of an ensemble of qubits to a common bosonic mode, can enable the rapid generation of non-Gaussian entangled spin-boson states without fine tuning of system parameters or initial conditions. However, the complexity of these states means that unlocking their utility for quantum-enhanced sensing with standard protocols would require the measurement of complex or typically inaccessible observables. To address this challenge, we develop a sensing scheme based on interaction-based readout that enable us to implement near-optimal quantum-enhanced metrology of global spin rotations or bosonic dipslacements using only spin measurements.  We show that our approach is robust to technical noise and imperfections and thus opens new opportunities to exploit complex entangled states generated by chaotic dynamics in current quantum science platforms such as trapped-ion and cavity-QED experiments. 
\end{abstract}

\maketitle

\noindent{\it Introduction}: 
Non-classical phenomena such as entanglement and quantum correlations present significant opportunities for the development of state-of-the-art sensors with performance far exceeding that possible with classical resources. However, the inherent fragility of entangled states remains a significant obstacle to practical applications \cite{Degen2017Quantumsensingreview,Toth2014Review,Pezze2018QuantumReview,Huelga1997Improvement}. The simplest solution is to ensure entanglement is generated and exploited on timescales much faster than those associated with intrinsic decoherence and technical noise. 
This has naturally motivated technical refinements in experiments to improve the ratio of coherent to dissipative timescales, as well as theory and experimental efforts to develop novel dynamical protocols that enhance the rate of entanglement generation \cite{GueryOdelin2019ShortcutsReview,JuliaDiaz2012Fast,Opatrny2016Counterdiabatic,Schloss2016Nonadiabatic,Song2016Generation, Pang2017Optimal, Cohn2018Bang, Palmero2019Towards} including the emulation of resonant pair production \cite{Sundar2023Bosonic,Barberena2022Fast} and saddle-point scrambling to create non-Gaussian states \cite{Muessel2015Twist,Mirkhalaf2018Robustifying,Pezze2018QuantumReview,Li2023Improving, Munoz2023Phase,Zhang2024Fast} in collective spin systems. 

In this light, many-body chaos is a natural candidate for accelerating entanglement generation for metrology~\footnote{We note that the distinct use of chaotic dynamics during interrogation (i.e., parameter encoding) has been proposed recently in the context of the kicked top \cite{Fiderer2018Quantum, Schuff2020Improving} and driven Bose-Josephson systems \cite{Liu2021Quantum}.}. For example, the quantum Fisher information (QFI), which quantifies the  potential of a quantum state for metrology, is known to generically grow exponentially fast in chaotic systems with a well-defined semiclassical limit \cite{LewisSwan2019Scrambling, Lerose2020Bridging} and saturate to large values \cite{Shi2024chaos}. The former feature can be traced back to the exponential divergence of initially adjacent trajectories of the associated chaotic classical model, which drives the rapid evolution of quantum fluctuations and thus QFI. 
However, QFI is a blunt metric as it presents the metrological potential of a quantum state optimized over all possible measurement strategies, regardless of the possible technical complexity. In fact, it is widely appreciated that the complexity of entangled states and the measurements required to fully exhaust their metrological potential are generally correlated \cite{Bollinger1996Optimal,Pezze2018QuantumReview,Helmut2014Fisher, Bohnet2016Quantum}. This suggests a significant challenge for the utilization of ``generic'' entangled states generated by chaotic dynamics, as they typically feature sophisticated structure (i.e., quickly evolve into non-Gaussian states) that requires the measurement of complex, many-body observables to access their full potential for metrology applications.

In this manuscript, we show that chaotic dynamics can nevertheless be a powerful and practical resource for quantum-enhanced metrology when combined with interaction-based readout (IBR). As demonstrated in recent theoretical and experimental works in the context of integrable systems \cite{Hosten2016Quantum,Macri2016Loschmidt,Davis2016Approaching,Linnemann2016Quantum,Nolan2017Optimal,Anders2018Phase,Schulte2020Ramsey,Gilmore2021Quantum,Colombo2022Time,Agarwal2022Quantifying,Li2023Improving}, an IBR protocol consists of incorporating an additional period of many-body dynamics after interrogation but before making a projective measurement on the probe. The general utility of IBR lies in the simplification of the measurement observable required to achieve the sensitivity predicted by the QFI, as well as favourable robustness to technical issues such as detection noise \cite{Hosten2016Quantum,Davis2016Approaching}. To provide a concrete example, we explore the Dicke model \cite{Dicke1954Coherence,Garraway2010Dicke,Kirton2019Introduction}, a paradigmatic example of classical and quantum chaos \cite{Emary2003Chaos,Emary2003Quantum,Perez2011Excited,Altland2012Quantum,Chavez2016Classical,Gietka2019Multipartite}, which describes many spins identically coupled to a single bosonic mode. The collective symmetry of the model enables efficient numerical simulation, either via exact or semiclassical methods, of relatively large ensembles of spins relevant for current and near-term experiments. Examples include mature quantum simulation platforms such as trapped ion arrays \cite{Safavi2018Verification,Cohn2018Bang} and cavity-QED \cite{Baumann2010Dicke,Klinder2015Dynamical,Zhang2018Dicke,Kroeze2018Spinor}. In contrast to prior works exploiting IBR, which primarily focused on integrable models of a collective spin, the combination of both spin and boson degrees of freedom in a chaotic model leads to additional complexity in designing optimal IBR protocols. An example is the variable degree to which the spin or boson degree of freedom can be accessed in different physical platforms, which motivates us to investigate appropriate IBR sequences and operating regimes that yield quantum-enhanced sensitivity with accessible and simple observables \cite{Lewisswan2024Exploiting}. On the other hand, we also demonstrate that this complexity provides opportunities to create ``all-purpose'' entangled quantum states that are simultaneously useful for quantum-enhanced sensing involving either spin or boson degrees of freedom without any fine tuning or additional manipulation. Our work can thus be relevant for a range of sensing applications, including timekeeping and magnetometry based on (global) qubit rotations \cite{Robinson2024clock,colombo2022clocks,muessel2014magnetometry}, and sensing weak forces or electric fields via coherent displacements of a bosonic mode \cite{Gilmore2021Quantum,Penasa2016microwave}. Contrary to na\"{\i}ve intuition based on the inherently erratic dynamics generated by classical chaos, we also demonstrate that our IBR approach is robust against typical sources of decoherence and noise, including perturbations away from the ideal IBR protocol.

\begin{figure}[bt]
\begin{center}
\includegraphics[width=1\columnwidth]{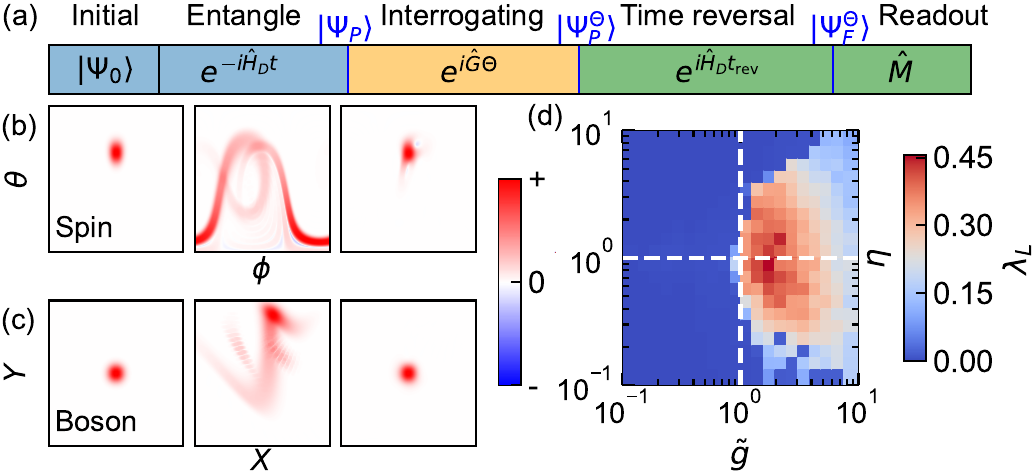}
\caption{
(a) Sensing protocol including time-reversal IBR. A complex probe state $\ket{\Psi_P} = e^{-i\hat{H}_D t}\ket{\Psi_0}$ is prepared through chaotic dynamics and then a parameter $\Theta$ is imprinted as $\ket{\Psi_P^{\Theta}} = e^{i\Theta\hat{G}}\ket{\Psi_P}$. The chaotic dynamics are then reversed to yield a relative simple final state $\ket{\Psi_f^{\Theta}} = e^{i\hat{H}_D t_{\mathrm{rev}}}\ket{\Psi_P^{\Theta}}$ for readout. 
(b) and (c) Phase-space illustration of the spin and boson degrees of freedom (via the associated Wigner distribution) for the initial state $\ket{\Psi_0}$, probe state $e^{-i\hat{H}_D t}\ket{\Psi_0}$ and final state $\ket{\Psi_F^\Theta}$. The example uses $(\theta_0,\phi_0,\alpha_0)=(1,0,0)$ for the initial state, generator $\hat G=\hat S_{\rm opt}$ for the interrogation and $N=50$. The central and right panels use an evolution time $gt=gt_{\rm rev}=4$ with parameters $\tilde g=2$ and $\eta=1$. (d) Lyapunov exponent $\lambda_L$ of the associated classical Dicke model. Dashed lines indicate $\tilde g=1$ and $\eta=1$ for reference.
} 
\label{fig:schematic}
\end{center}
\end{figure}

{\it Sensing protocol:}
A schematic outline of our sensing protocol is presented in Fig.~\ref{fig:schematic}(a). We start by preparing an uncorrelated product state $\ket{\Psi_0}$ of a single boson mode and $N$ qubits, where the latter is assumed to be prepared collectively (i.e., equivalently characterized by a single large spin of length $N/2$). Each degree of freedom is assumed to be in a coherent state with isotropic quantum fluctuations (projection noise). We visualize this in the leftmost panels of (b) and (c) in terms of the corresponding Wigner function, which is a representation of the quantum state in phase space spanned by effective position and momentum quadratures, $(X,Y)$, for the boson mode and angular co-ordinates, $(\theta,\phi)$, that parameterize the collective spin on the associated collective Bloch sphere \cite{Koczor2020Fast,QuTiP2}. 
Quantum evolution under the Dicke Hamiltonian $\hat H_D$ in a parameter regime featuring classically chaotic dynamics [see panel (d) and later discussion] rapidly reshapes the quantum noise of the spin-boson state into a complex, erratic pattern in phase space [middle panels of (b) and (c)]. 
The fluctuations of the quantum state intrinsically relate to its utility for metrology via the QFI, $F(\hat G)=4{\rm Var}(\hat G)$ where $\hat G$ is the generator of the perturbation $e^{i\Theta\hat G}$ that imprints an unknown parameter $\Theta$ on the probe state $\ket{\Psi_P}=e^{-iH_{D}t}\ket{\Psi_0}$. The QFI defines the quantum Cramer-Rao bound (QCRB) $(1/\Delta \Theta)^2\leq F(\hat G)$ on the minimum uncertainty in the estimation of the parameter $\Theta$.
The delocalization and associated build-up of intricate structure \cite{Zurek2001SubPlanck}, including small fringes and negative regions associated with non-classicality \cite{WallsMilburnBook}, in the representation of the probe state in phase space are desirable features for metrology~\footnote{Note that our distributions actually underestimates the true complexity and structure of the quantum state as in each case it is obtained by integrating out the complementary degree of freedom}. 
Specifically, a small perturbation of the probe state in phase space --- such as those generated by spin rotations or bosonic displacements --- cause it to rapidly become distinguishable (i.e., orthogonal) to the unperturbed one \cite{Zurek2001SubPlanck,Toscano2006Subplanck}, leading to a large QFI \cite{Braunstein1994qfi}. In particular, the QFI will be increased relative to the initial state $\ket{\Psi_0}$ (left panels), whose metrological potential is comparatively limited as a result of the characteristic quasiclassical isotropic fluctuations.

The utility of states generated by chaos is enriched relative to other common approaches that generate, e.g., (over-)squeezed and cat states, as their complex fine structure generically makes them sensitive to perturbations along an arbitrary rotation axis or displacement direction and eliminates the need for additional re-orientation or characterization \cite{Holland2023optimalgenerator,Shi2024chaos}. However, this same structure results in challenges exploiting these states for metrology, as they cannot be characterized by simple measurement observables. To overcome this challenge we harness IBR in the form of a time-reversal echo, wherein the initial dynamics is run in reverse by inverting the sign of the Hamiltonian. In the absence of any perturbation by a spin rotation or bosonic displacement, the quantum state will return to the initial condition $\ket{\Psi^{\Theta=0}_F} = e^{iH_{D}t}\ket{\Psi_P} = \ket{\Psi_0}$. However, a small perturbation of the probe state leads to the final state $\ket{\Psi^{\Theta}_F} = e^{iH_{D}t}e^{i\Theta\hat{G}}\ket{\Psi_P}$ deviating away from the initial condition. Parameter estimation can then be efficiently carried out by making a discriminating measurement that distinguishes the return to the initial uncorrelated product state from other scenarios. 

{\it Model:}
As a concrete example, we consider the Dicke model for $N$ qubits \cite{Dicke1954Coherence} described by the Hamiltonian,
\begin{equation}\label{eq:dickemodel}
    \hat H_D=-\frac{2g}{\sqrt{N}}(\hat a+\hat a^{\dagger})\hat S_x+\delta\hat a^{\dagger}\hat a+\Omega \hat S_{z}\,,
\end{equation}
where we have set $\hbar = 1$. Here, $\hat a^{\dagger}$ ($\hat a$) (annihilates) a boson with energy $\delta$, and we have used collective spin operators $\hat S^\alpha=1/2\sum_{j=1}^{N}\hat \sigma_j^\alpha$ defined in terms of Pauli operators for spin $j$, $\hat\sigma_j^\alpha$ ($\alpha=x$, $y$, and $z$). The relative splitting of the qubit states is set by $\Omega$, and $g$ characterizes the spin-boson coupling. For reference, we introduce the ratio $\eta\equiv\Omega/\delta$ that loosely characterizes the relative contribution of the two degrees of freedoms to the dynamics \cite{Lewisswan2021Characterizing}: For $\eta<<1$, the dynamics of Hamiltonian \eqref{eq:dickemodel} is dominated by the collective spin and reduces to an LMG model \cite{LMG1965,SafaviNaini2017dicke}. On the other hand, for $\eta>>1$, the spin dynamics is slaved to the dominant bosons \cite{Lewisswan2021Characterizing}. We additionally introduce the rescaled coupling $\tilde g\equiv g/g_c$ where $g_c=\frac{\sqrt{\delta\Omega}}{2}$ is the critical coupling for the ground-state quantum phase transition in the model \cite{Emary2003Chaos}. 

\begin{figure}[bt]
\begin{center}
\includegraphics[width=1\columnwidth]{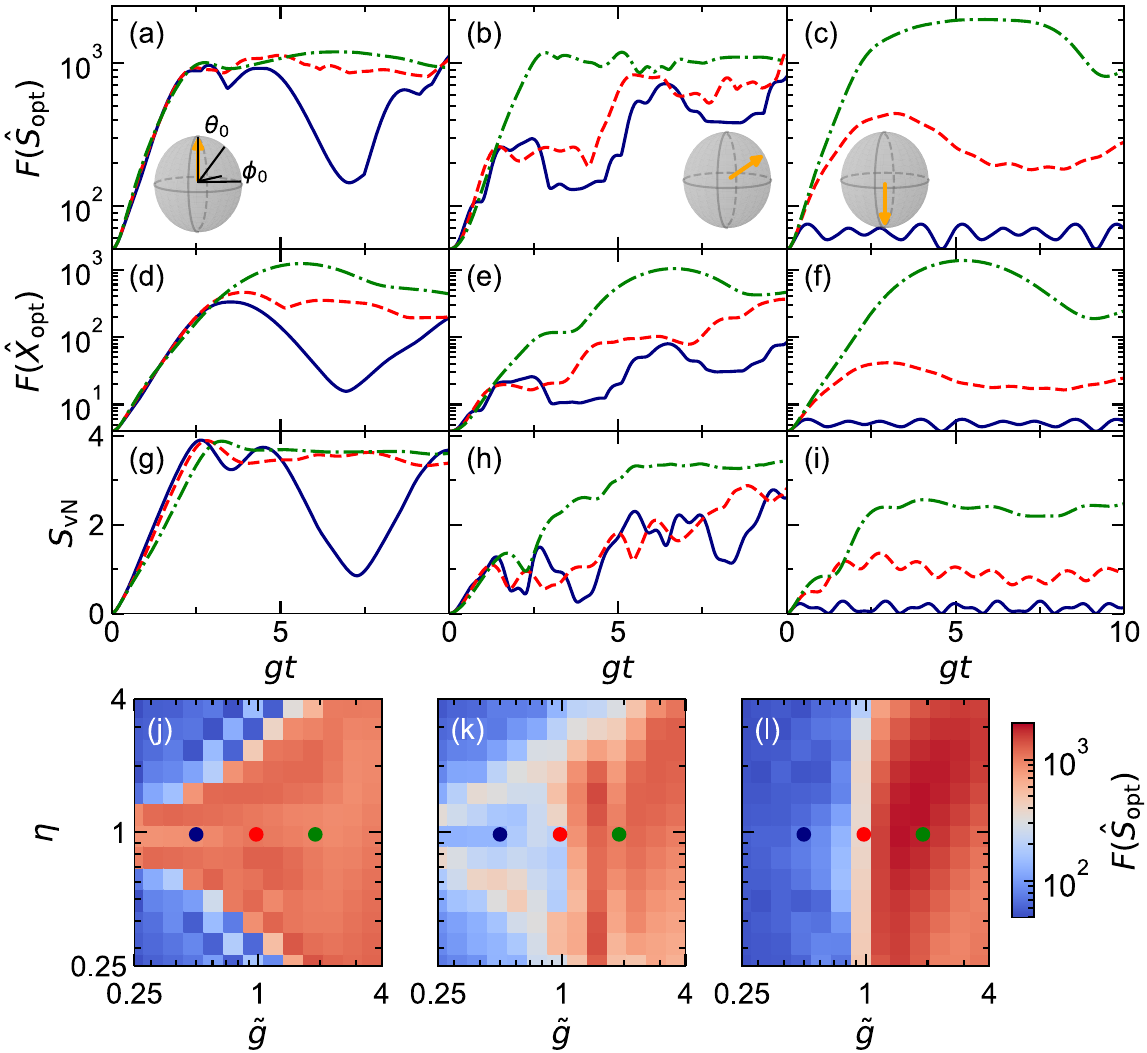}
\caption{
(a)--(c) Evolution of optimal QFI for spin rotations, $F(\hat S_{\rm opt})$. We compare dynamics initiated at classical fixed points (a) $(\theta_0,\phi_0,\alpha_0)=(0,0,0)$ and (c) $(\pi,0,0)$ with a generic state (b) $(1,0,0)$. All panels show data for $\eta=1$, and $\tilde g=0.512$ (blue solid lines), 1 (red dashed lines), and 1.953 (green dashed dotted lines), with $N=50$. (d)--(f) Evolution of optimal QFI for bosonic displacements, $F(\hat X_{\rm opt})$. Initial conditions and parameters are identical to (a)--(c). (g)--(i) Spin-boson entanglement entropy $S_{\rm vN}$ corresponding to timetraces in (a)--(c) and (d)--(f). (j)--(l) Parameter dependence of $F(\hat S_{\rm opt})$ at fixed $gt=4$ for the initial states of (a)--(c). Colored dots indicate values of $(\tilde{g},\eta)$ used for curves in upper panels.
} 
\label{fig:qfi}
\end{center}
\end{figure}

We consider an initial state $\ket{\Psi_0}=\ket{\theta_0,\phi_0\rangle\otimes|\alpha_0}$, where $\ket{\theta_0,\phi_0}$ is a spin coherent state pointing along the $(\theta_0,\phi_0)$ directions on the Bloch sphere [see e.g., inset of Fig.~\ref{fig:qfi}(a)], and $\ket{\alpha_0}$ is a bosonic coherent state such that $\langle \hat{a}^{\dagger} \hat{a} \rangle = \vert \alpha_0 \vert^2$. Without loss of generality, we set $\alpha_0=0$ (i.e., a vacuum state) in this manuscript and primarily focus on an initial spin orientation defined by $\theta_0=1$, $\phi_0=0$. 

To first identify regimes of classical chaos, we explore the dynamical phase diagram of the associated classical dynamics (obtained by replacing operators $\hat{O}$ with $c$-number phase space variables $\mathcal{O}$ in the above Hamiltonian) in terms of the Lyapunov exponent $\lambda_L$ \cite{Chavez2016Classical,LewisSwan2019Scrambling,Skokos2010Book}. Figure \ref{fig:schematic}(d) shows $\lambda_L$ as a function of $\tilde g$ and $\eta$ for our chosen initial state. Classical chaos is signalled by $\lambda_L > 0$, which indicates an exponential divergence of initially adjacent classical trajectories as a function of time. We thus identify a chaotic regime that emerges when $\tilde{g} \gtrsim 1$ and $\eta\sim 1$ (white dashed lines). 

{\it Quantum Fisher information:}
To demonstrate the utility of the underlying classical chaos for metrology, we similarly map out a phase diagram for the quantum dynamics in terms the QFI as a function of $\tilde{g}$ and $\eta$. Note that all results for the quantum dynamics in this manuscript are obtained via efficient integration of the Schrödinger equation using the Krylov subspace method \cite{IterativeMethodBook2023} and with a truncated Fock basis. 
For generality, we consider the QFI for two generators corresponding to distinct sensing tasks: First, a phase shift imprinted via a global spin rotation about an axis $\hat n_s=(\theta_s, \phi_s)$ defined by elevation and azimuthal angles $\theta_s$ and $\phi_s$, respectively, such that $\hat G=\hat S_{\hat n_s}$. Secondly, we consider displacements of the boson mode along a direction defined by an angle $\varphi$ such that $\hat G=\hat X_{\varphi}=\hat a^{\dagger}e^{i\varphi}+\hat a e^{-i\varphi}$.

In Fig.~\ref{fig:qfi}(a)--(c) we plot the time evolution of the maximal QFI $F(\hat S_{\rm opt})$, obtained by optimizing the axis of the spin rotation at each time.
Panel (b) shows results for the initial state $(\theta_0,\phi_0,\alpha_0)=(1,0,0)$ with $N = 50$ (same as Fig.~\ref{fig:schematic}).
We focus on dynamics emblematic of the regular (i.e., $\lambda_L = 0$) and chaotic ($\lambda_L > 0$) regimes by fixing $\eta=1$ and choosing $\tilde g=0.512$ (blue solid lines) and $1.953$ (green dashed dotted lines), respectively. We additionally show a case at the boundary of these regimes with $\tilde{g} = 1$ (red solid lines). 
The chaotic example displays an initial exponential growth of the QFI before saturating to a value $\propto N^2$ (the QFI is bounded by $F_Q \leq N^2$ for spin rotations) at $t\sim \log(N)$ \cite{Gietka2019Multipartite, LewisSwan2019Scrambling, Lerose2020Bridging}. On the contrary, in the absence of chaos the QFI grows much more slowly and reaches a similar maximum value at much longer times. The difference in timescales is expected to be more pronounced as $N$, and thus the maximum attainable QFI, is increased as the QFI in the chaotic regime will evolve exponentially over multiple decades in time according to the Lyapunov exponent $F(\hat S_{\rm opt})\sim e^{2\lambda_L t}$ until saturation \cite{LewisSwan2019Scrambling,SM}. For our relatively small system, we can illustrate the distinction between the growth of QFI in the chaotic and regular regimes by plotting $F(\hat S_{\rm opt})$ at fixed $gt=4$ [$\approx\log(N)$, the scrambling time in the chaotic regime] in Fig.~\ref{fig:qfi}(k). We observe that large values of the QFI correlate with the regions of chaos ($\lambda_L \neq 0$) in Fig.~\ref{fig:schematic}(d). We interpret this result as demonstrating that the QFI has already saturated in the chaotic regime, whereas in the regular regime the QFI is still undergoing slow growth.

The metrological utility of the entangled spin-boson state is not limited to spin rotations. In Fig.~\ref{fig:qfi}(d)--(f) we show that the QFI associated with bosonic displacements simultaneously features rapid growth in the chaotic regime and saturates near the relative HL (set by the average boson occupation $\langle \hat{a}^{\dagger}\hat{a} \rangle$). For this data we use the same initial state as Fig.~\ref{fig:qfi}(a)--(c) and construct the QFI $F(\hat X_{\rm opt})$ where $\hat G=\hat X_{\rm opt}$ is chosen by optimizing $\varphi$ to maximize the QFI.

We observe that a key benefit of chaotic dynamics is the relative ubiquity of the exponentially fast generation of metrological useful entanglement without fine tuning of initial conditions or other parameters. In Ref.~\cite{SM} we demonstrate that the fast growth and late-time saturation of the QFI is largely insensitive to the choice of spin rotation axis or boson displacement direction, a result which is consistent with Fig.~\ref{fig:schematic} but also recent work in Ref.~\cite{Shi2024chaos} that predicts chaotic systems should typically feature isotropic fine structure in phase space. We similarly observe that the fast growth of entanglement is not tied to a specific initial condition, which contrasts with approaches using integrable dynamics that leverage classical saddle points to gain a similar speed-up \cite{LewisSwan2019Scrambling, Li2023Improving,Sundar2023Bosonic}. To underline this, we compare to the dynamics of the QFI for the special initial states $(\theta_0,\phi,\alpha_0) = (0,0,0)$ and $(\pi,0,0)$, which correspond to a pair of classical fixed points of the Dicke model.  The QFI for both spin rotations and boson displacements are shown in panels (a), (d) and (c), (f) of Fig.~\ref{fig:qfi}, respectively. 

We observe the rate of growth of the QFI and maximum value attained with the ``generic" initial state of $(\theta_0,\phi_0,\alpha_0) = (1,0,0)$ is similar to those for the fixed points in the chaotic regime, $\tilde{g} > 1$. This demonstrates that it is not necessary to fine tune the initial condition to generate metrologically useful states, nor is there a substantial quantitative advantage to be found in terms of the growth rate or saturation value of the QFI. On the other hand, in the regular regime, $\tilde{g} < 1$, only the fixed 
point $(\theta_0,\phi_0,\alpha_0) = (0,0,0)$ generates a large QFI. In this case, the initial condition is an unstable classical fixed point and the metrological utility has previously been studied through the lens of effective two-mode squeezing \cite{Barberena2024Fast}. For further characterization, we systematically map out $F(\hat S_{\rm opt})$ as a function of $\tilde g$ and $\eta$ at $gt=4$ for the fixed points in Fig.~\ref{fig:qfi} (j) and (l). This reveals that in the regime $\tilde{g} < 1$ the large QFI obtained with the unstable fixed point is confined to $\eta \approx 1$, i.e., particular combinations of parameters such that $\delta \approx \Omega$. In Ref.~\cite{SM} we show similar results for the QFI with other $\theta_0$ and finite $\alpha_0 \neq 0$, which demonstrates the generality of our findings for generic initial conditions in the chaotic regime.

{\it Impact of spin-boson entanglement:}
It is intuitive to suspect that the complex, delocalized entangled states that are generated by chaotic dynamics, as typified by the results shown in Fig.~\ref{fig:schematic}, will require sophisticated, technically challenging measurements of high-order correlations or many-body observables to realize quantum-enhanced parameter estimation at the level promised by the QFI. Indeed, this difficulty in utilizing generic non-Gaussian quantum states for metrology is widely appreciated \cite{Strobel2014Fisher,Bohnet2016Quantum,Colombo2022Time}. However, an important distinction from previous investigations of non-Gaussian states, which have typically focused on systems involving only a collective spin, is the additional level of complexity that is introduced by the interplay and entanglement of the spin and boson degrees of freedom in the Dicke model. 

We illustrate this in Fig.~\ref{fig:qfi}(g)--(i) by showing the spin-boson entanglement entropy $S_{\rm vN}=-\Tr \rho_{\rm sp}\log \rho_{\rm sp}$, where $\rho_{\rm sp}$ is the reduced density matrix describing the spin degreee of freedom. Comparing to the prior panels, we observe that the growth of $F(\hat S_{\rm opt})$ and $F(\hat X_{\rm opt})$ is accompanied by a commensurate increase in the entanglement between the two degrees of freedoms. Difficulty arises as this entanglement implies that saturating the QCRB set by the QFI will require the measurement of correlations between spin and boson observables, which can be technically challenging. For example, in trapped ion arrays only spin observables are typically directly accessible with sufficient resolution and fidelity. Conversely, the entanglement dictates that each degree of freedom features excess projection noise when considered independently, which limits the achievable sensitivity when measuring spin or boson observables separately \cite{Lewisswan2024Exploiting}. 

\begin{figure}[bt]
\begin{center}
\includegraphics[width=1\columnwidth]{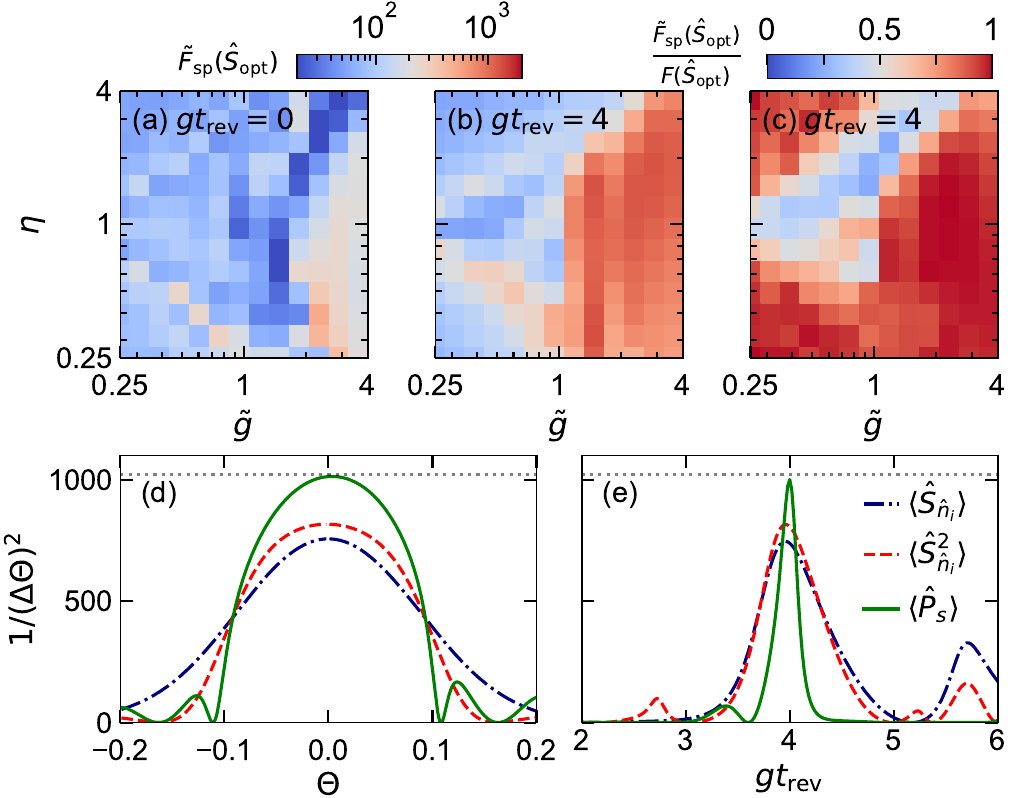}
\caption{Spin QFI, $\tilde F_{\rm sp}(\hat S_{\rm opt})$ (a) immediately after interrogation and (b) after subsequent time-reversal with $t_{\rm rev}=t$, as a function of $\tilde{g}$ and $\eta$. Panel (c) shows the normalized spin QFI $\tilde F_{\rm sp}(\hat S_{\rm opt})/F(\hat S_{\rm opt})$ after time-reversal [i.e., same data as (b)]. All panels use $(\theta_0,\phi_0,\alpha_0)=(1,0,0)$, $gt = 4$ and $N = 50$. (d) Inverse sensitivity $1/(\Delta\Theta)^2$ as a function of the parameter $\Theta$ for different measurement observables [see legend in (e)]. 
(e) Dependence of inverse sensitivity on reversal time $t_{\rm rev}$. Data is obtained at fixed $\Theta=1/N$ with parameters $\tilde g=1.953$ and $\eta=1$. Legend indicates the measurements $\hat M$ used to estimate $\Theta$. Dotted horizontal line in (d) and (e) indicates the QCRB defined by the QFI of the full system.
} 
\label{fig:spinqfi}
\end{center}
\end{figure}

To demonstrate this issue we compute the spin QFI $\tilde F_{\rm sp}(\hat S_{\rm opt})$ associated with the state $\rho_{\rm sp}$ of the spins. One can understand this QFI as defining an alternative bound on the achievable sensitivity, $\Delta\Theta \geq \tilde F_{\rm sp}^{-1}$, when constrained to only measurements of the spin degree of freedom (on the other hand, the QFI $F(\hat S_{\rm opt})$ computed with respect to the full spin-boson state yields a bound on the sensitivity after optimizing over \emph{all} possible measurements involving both degrees of freedom). Figure \ref{fig:spinqfi}(a) shows $\tilde F_{\rm sp}(\hat S_{\rm opt})$ as a function of $\tilde{g}$ and $\eta$ and should be contrasted with the QFI in Fig.~\ref{fig:qfi}(k). We observe that $\tilde F_{\rm sp}(\hat S_{\rm opt})$ is notably reduced by an order of magnitude compared to the full QFI, particularly in the chaotic regime ($\tilde{g} > 1$) where the spin-boson entanglement is also largest.

{\it Time-reversal IBR:}
To overcome this challenge we utilize an IBR protocol to instead map the complex quantum state after the interrogation to a simple final state that ideally encodes the displacement or rotation parameter in a single degree of freedom. As an example, we consider a time-reversal IBR, where the initial entangling dynamics used to prepare the probe state is undone by evolving ``backwards'' with $\hat H'_D=-\hat H_D$ for a duration $t_{\rm rev}$. 
In an experiment, flipping the sign of the Hamiltonian can be achieved by directly controlling the sign of each parameter $g$, $\delta$ and $\Omega$, or global spin rotations.  
The utility of the time-reversal IBR is demonstrated in panels Fig.~\ref{fig:spinqfi}(b), which shows the spin QFI $\tilde F_{\rm sp}(\hat S_{\rm opt})$ for the balanced case $t_{\rm rev}=t$. Additionally, Fig.~\ref{fig:spinqfi}(c) shows the ratio $\tilde F_{\rm sp}(\hat S_{\rm opt})/F(\hat S_{\rm opt})$ at $t_{\rm rev}=t$. Comparing to Fig.~\ref{fig:spinqfi}(a), we observe that the spin QFI is dramatically increased. In particular, the spin QFI nearly saturates the QFI of the full system in the (metrologically benefical) chaotic regime and when $\eta \sim 1$ (see panel (c), where the ratio approaches one), indicating that detection of the spin state can be sufficient to nearly saturate the QCRB.  Similar results are obtained for the equivalent boson QFI, $\tilde{F}_b$, and we show these in Ref.~\cite{SM}. Typically, we observe that spin (boson) measurements become optimal when $\eta < 1$ ($\eta > 1)$, which is consistent with the dynamics becoming dominated by the spin (boson) degree of freedom \cite{SM}. We note that, in principle, if the spin QFI is small but there is a large boson QFI, this can still be exploited for metrology with spin-only readout by supplementing the time-reversal IBR with a collective sideband coupling \cite{SM} to transfer information between the bosons and spins as proposed recently in Ref.~\cite{Lewisswan2024Exploiting}. 
Overall, our results illustrate that time-reversal can efficiently disentangle the system and approximately map information about the encoded parameter $\Theta$ for efficient estimation via measurement of a single degree of freedom.

In general, the sensitivity attainable with a measurement of an observable $\hat{M}$ is given by $(\Delta\Theta)^2 = \mathrm{Var}(\hat{M})/\vert \partial_{\Theta}\langle \hat{M} \rangle\vert^2$, which when $\hat{M}$ is limited to spin observables is bounded by $(\Delta\Theta)^2 \geq \tilde F_{\rm sp}^{-1}$
The optimal spin observable that saturates the equality for an initial spin-boson product state and assuming $t_{\rm rev}=t$ is the projection into the initial, uncorrelated spin state: $\hat{P}_s = \ket{\theta_0,\phi_0}\bra{\theta_0,\phi_0}$ \cite{SM}. Such a measurement is feasible in many platforms, although it requires the technical capability to discriminate single spin flips (see later discussion and Ref.~\cite{SM}). However, enhanced sensitivity can also be achieved with simpler observables, including moments of the associated global spin projection $\hat S^m_{\hat n_i}$ where the axis $\hat n_i$ matches the polarization of the initial state $|\theta_0,\phi_0\rangle$.

In Fig.~\ref{fig:spinqfi}(d), we compare the inverse sensitivity $1/(\Delta\Theta)^2$ for $\hat M=\hat S_{\hat n_i}$ (blue dashed dotted line) and $\hat S^2_{\hat n_i}$ (red dashed line) with the ideal case $\hat P_s$ (green solid line, equivalent to $F_{\rm sp}$) for $\tilde{g} = 1.953$ and $gt_{\rm rev} = gt = 4$ as a function of the spin rotation angle $\Theta$. As expected for a time-reversal IBR, the optimal sensitivity identically occurs for $\Theta \to 0$. Importantly, we observe that the best achievable (inverse) sensitivity with $\hat S_{\hat n_i}$ is only reduced by about $30\%$ from that obtained with $\hat P_s$ but features a larger dynamic range (i.e., window of $\Theta$ with sensitivity beyond the SQL) \cite{davis2017advantages}. The measurement of higher moments, e.g., $\hat{S}^2_{\hat{n}_i}$, improves the sensitivity although this is expected to be accompanied by an increase in fragility to detection noise \cite{Guan2021Tailored}. In Fig.~\ref{fig:spinqfi}(e), we additionally show the dependence of the optimal sensitivity (i.e., $\Theta \to 0$) as a function of reversal time $t_{\rm rev}$ for fixed $gt = 4$. The optimal sensitivity is achieved for the balanced case $t_{\rm rev} = t$, although the window of $t_{\rm rev}$ featuring enhanced sensitivity is highly dependent on the observable (e.g., $\hat{S}_{n_i}$ shows a broader feature than $\hat{P}_s$) and the working value of $\Theta$.

{\it Robustness to technical imperfections:}
Given the complexity of the states generated by chaotic dynamics and thus our reliance on IBR for practical readout, it is important to characterize the robustness of our protocol to imperfections. In particular, we study the fragility of our time-reversal IBR to small parameter fluctuations, as well as the sensitivity of our results to classical noise in the initial state preparation. 

As an illustrative example of the former, we consider a scenario where the oscillator frequency $\delta$, which is typically a detuning relative to some reference, is poorly calibrated such that the sign cannot be precisely flipped for perfect time reversal. Specifically, we consider an imperfect time-reversal IBR with $\hat H'_D$ characterized by quenching $g \to -g$, $\Omega \to -\Omega$, $\delta \to \delta_{\rm rev} \neq -\delta$ such that $\hat{H}'_D \neq -\hat{H}_D$ in general. This scenario is relevant to, e.g., recent work \cite{Gilmore2021Quantum} using large $2$D ion arrays in a Penning trap for displacement sensing, wherein frequency fluctuations were found to be a limiting factor for performance. To be concrete, we treat $\delta_{\rm rev}$ as a Gaussian random variable that changes from shot-to-shot with mean $-\delta$ and variance $\sigma_{\delta}^2$, and compute the sensitivity from the averaged signal accordingly. 

\begin{figure}[bt]
\begin{center}
\includegraphics[width=1\columnwidth]{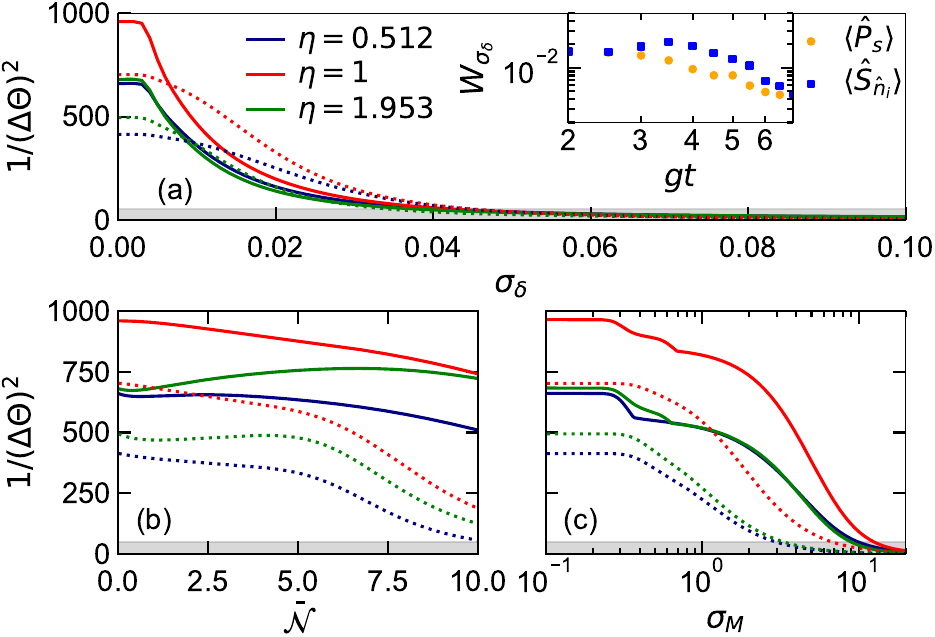}
\caption{
(a) Robustness of inverse sensitivity $1/(\Delta \Theta)^2$ as a function of rms fluctuations $\sigma_\delta$ in boson frequency during time-reversal (see text). Inset: Full-width-at-half-maximum $W_{\delta}$ (relative to sensitivity at $\sigma_\delta=0$)
as a function of $t$. (b) Dependence of $1/(\Delta \Theta)^2$ on initial thermal boson occupation $\cal N$. Solid (dotted) lines in (a) and (b) use a measurement of $\hat M=\hat P_s$ ($\hat M=\hat S_{\hat n_i}$). (c) Robustness of sensitivity to detection noise characterized by $\sigma_M$ (see text). Dotted lines are obtained from $\hat M=\hat S_{\hat n_i}$, while solid lines are computed using the classical Fisher information $F_C$ (computed with respect to an optimal spin projection basis \cite{SM}) which is related to the inverse sensitivity as $1/(\Delta\Theta)^2 = F_C$.
All panels use $\tilde g=1.953$, $t_{\mathrm{rev}} = t$ and $N=50$. Values of $\eta$ are indicated in the legend of (a) [except for the inset of (a) which uses $\eta = 1$].
} 
\label{fig:noise}
\end{center}
\end{figure}

Figure \ref{fig:noise}(a) shows the inverse sensitivity $1/(\Delta \Theta)^2$ for a measurement of $\hat P_s$ (solid lines) and $\hat S_{\hat n_i}$ (dotted lines) in the chaotic regime, $\tilde g=1.953$, $gt = gt_{\rm rev} = 4$ and different values of $\eta$ (we fix $\delta = 1$ and vary $\Omega$ and $g$). At resonance ($\eta=1$, red curves), we observe a robustness (i.e., $1/(\Delta \Theta)^2$ is effectively unchanged) to very small frequency fluctuations $\sigma_{\delta} < 0.01$, but overall quantum-enhancement is eventually lost when $\sigma_{\delta} \gtrsim 0.03$. We note that the sensitivity obtained with the mean spin projection $\hat S_{\hat n_i}$ is typically more robust than that for $\hat P_s$. Quantifying the tolerance as $W_{\sigma_\delta}$, which is defined as the value of $\sigma_\delta$ at which the sensitivity $1/(\Delta \Theta)^2$ drops to half of the $\sigma_\delta=0$ value, we show the tolerance to frequency fluctuations as a function of $t$ (fixing $t_{\rm rev}=t$) in the inset of Fig.~\ref{fig:noise}(a). We observe that $W_{\sigma_\delta}$ decreases at longer times, which is consistent with intuition that suggests the effects of the frequency fluctuations should eventually overwhelm the small perturbation created by the spin rotation or boson displacement. We note that the when converted to a relative tolerance of typically about $1$-$3\%$, the required stability in $\delta$ is consistent with values already achieved in Ref.~\cite{Gilmore2021Quantum}, but performance can be improved by operating with larger $\delta$ (and thus $g$ and $\Omega$). Our results are qualitatively similar for the other values of $\eta$ (see blue and green curves) plotted and thus suggest that time-reversal IBRs can be implemented sufficiently accurately to enable quantum-enhancement. 

Classical noise in the initial conditions can be introduced by imperfect state preparation, such as a residual thermal occupation $\bar {\cal N}$ of the boson mode after cooling for our example initial condition. Figure \ref{fig:noise}(b) shows $(1/\Delta\Theta)^2$ as a function of $\bar {\cal N}$ in the chaotic regime
with $(\theta_0,\phi_0) = (1,0)$ unchanged.
We find that the sensitivity attained with a measurement of $\hat P_s$ (solid lines) is very robust regardless of $\eta$, and in fact the sensitivity can even \emph{increase} with $\bar {\cal N}$. However, the sensitivity attained with $\hat{S}_{\hat{n}_i}$ (dotted lines) does show a drop off with increasing occupation, though it remains robust for about $\bar {\cal N}\lesssim 6$. This occupation coincides with that achieved in Ref.~\cite{Gilmore2021Quantum} using only Doppler cooling of the collective motion of the trapped ion array.

Lastly, a desirable trait of IBR protocols is their typical robustness to detection noise \cite{Hosten2016Quantum,Davis2016Approaching,Nolan2017Optimal,Mirkhalaf2018Robustifying}. To demonstrate this for our protocol, we consider Gaussian detection noise of characteristic strength $\sigma_M$ (see Ref.~\cite{SM} for details). 
Figure \ref{fig:noise}(c) shows $(1/\Delta\Theta)^2$ as a function of the noise strength $\sigma_M$ from a measurement of $\hat S_{\hat{n}_i}$ (dotted lines) as well as the classical Fisher information (CFI) $F_{C}$ obtained from the counting statistics of $\hat S_{\hat{n}_i}$ ($(1/\Delta\Theta)^2 = F_c^{-1}$, solid lines), and which is equivalent to our usual projective measurement $\hat{P}_s$ for $\sigma_M = 0$. In both cases, we optimize the measurement axis $\hat{n}_i$ for each $\sigma_M$. We find that saturating the QCRB via the CFI requires single-particle resolution ($\sigma_M \ll 1$), consistent with $\hat{P}_s$, but from a less restrictive perspective the CFI maintains quantum-enhanced sensitivity for detection resolution better than the projection noise ($\sigma_M \lesssim \sqrt{N}$). A simpler measurement of the mean spin projection $\hat S_{\hat{n}_i}$ is slightly less robust, but still yields quantum-enhanced sensitivity without single-particle resolution. 

{\it Conclusion:}
Our results suggest that chaotic quantum dynamics can be a resource for quantum-enhanced sensing without fine-tuned initial state preparation, choice of operating parameters or re-orientation for interrogation. This is enabled by a time-reversal IBR that minimizes the technical requirements for readout, and which is robust to common experimental imperfections. While our approach is broadly relevant for any quantum sensing platform where chaotic dynamics is accessible, our specific findings for the Dicke model can be directly applied in current trapped-ion and cavity-QED platforms. Realizations in these systems will also motivate a detailed, platform-specific analysis of decoherence. Future work may also explore how IBR protocols can be optimized in the presence of decoherence \cite{Juan2024qfiopt,MacLellan2024endtoend} or to maximize resilience to technical imperfections such as detection noise \cite{haine2024IBR}.

{\it Acknowledgements:} We acknowledge support by NSF through Grant No. PHY-2110052. Y.Z. acknowledge support from Dodge Family Postdoc Fellowship at the University of Oklahoma. The computing for this project was performed at the OU Supercomputing Center for Education \& Research (OSCER) at the University of Oklahoma (OU).

\bibliographystyle{apsrev4-1}
\bibliography{Reference}

\phantom{a}
\newpage
\setcounter{figure}{0}
\setcounter{equation}{0}
\setcounter{table}{0}

\renewcommand{\thetable}{S\arabic{table}}
\renewcommand{\thefigure}{S\arabic{figure}}
\renewcommand{\theequation}{S\arabic{equation}}
\renewcommand{\thepage}{S\arabic{page}}

\renewcommand{\thesection}{S\arabic{section}}

\onecolumngrid

\begin{center}
\setcounter{page}{1}
{\large \bf Supplemental Material:\\
Harnessing quantum chaos in spin-boson models for all-purpose quantum-enhanced sensing}
\vspace{0.3cm}
\end{center}
\vspace{0.6cm}

\twocolumngrid

\label{pagesupp}

\section{Quantum Fisher information of probe states}

The results of the main text primarily focus on a single initial condition $\ket{\Psi_0}=\ket{\theta_0,\phi_0}\otimes\ket{\alpha_0}$ with $(\theta_0,\phi_0,\alpha_0)=(1,0,0)$ for $N = 50$.
Moreover, the data presented in Figs.~2-4 uses spin rotations $\hat G=\hat S_{\hat n_s}$ and boson displacements $\hat X_{\varphi}$ that are optimized over all rotation or displacement axes ($\hat n_s$ and $\varphi$) to achieve the best QFI or related sensitivity for a given set of parameters, entangling times etc. Here, we provide additional results, including an illustration of the connection between QFI growth and chaotic dynamics in the semiclassical limit of large $N$, as well as the behaviour of the QFI and sensing performance for other choices of $\ket{\Psi_0}$ and $\hat G$.

\noindent{\bf Semiclassical limit --- }
For the exact dynamics shown in Fig.~2 of the main text, the expected exponential growth of the QFI is compromised by the finite system size ($N=50$). For larger $N$, we expect an even more pronounced difference in the QFI for regular and chaotic dynamics \cite{LewisSwan2019Scrambling}. We show this by using the semiclassical truncated Wigner approximation (TWA) method to treat a large system \citeSI{Polkovnikov2010Phase}. TWA incorporates the effects of quantum fluctuations at first-order beyond a classical (mean-field) treatment by: i) randomly sampling the initial conditions of spin and boson $c$-number variables $(S_x,S_y,S_z,\alpha)$ according to the Wigner function of the initial state \citeSI{OLSEN2009sampling} and then ii) integrating the mean-field (classical) equations of motion for each initial state. Quantum expectation values are obtained by appropriate averaging over the ensemble of classical trajectories.

\begin{figure}[bt]
\begin{center}
\includegraphics[width=1\columnwidth]{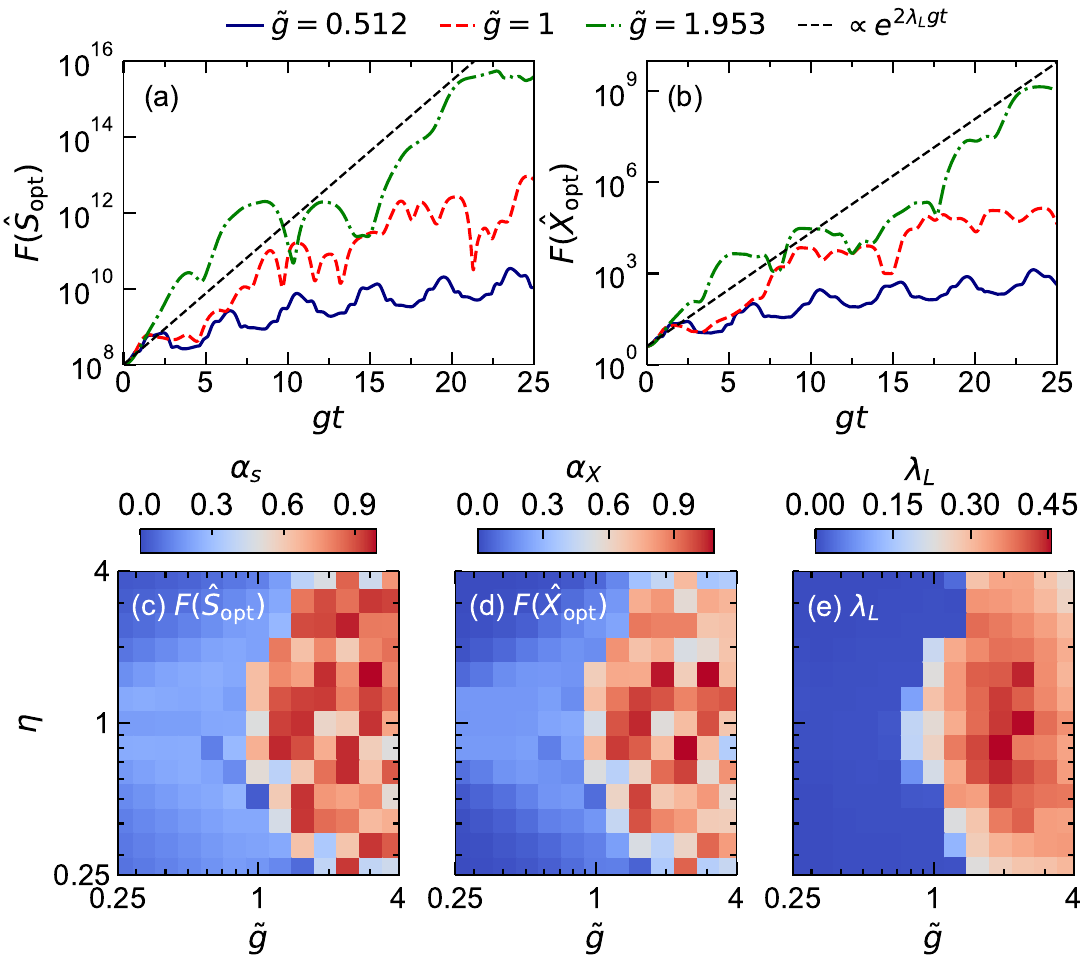}
\caption{{\it QFI in the semiclassical limit.} (a) Maximum QFI for spin rotations, $F(\hat S_{\rm opt})$, as a function of $gt$ (the spin rotation axis $\hat S_{\hat n_s}$ is optimized  at each time point). The initial state and Hamiltonian parameters are the same as Fig.~2(b) of the main text but we use $N=10^8$. Results are obtained with TWA (see text). Black dashed line indicates $Ne^{2\lambda_L gt}$ as a reference. (b) Same as (a) but for boson displacements, $F(\hat X_{\rm opt})$, (the displacement direction is optimized  at each time point). See Fig.~2(e) of the main text for comparison. Black dashed line indicates $4e^{2\lambda_L gt}$ as a reference.(c) and (d) Density plots of parameter $\alpha_s$ ($\alpha_X$) obtained by fitting the TWA results for $F(\hat S_{\rm opt})=A_se^{\alpha_s gt}$ [$F(\hat X_{\rm opt}) =A_Xe^{\alpha_X gt}$] for different $\tilde g$ and $\eta$ with $N = 10^8$. We only fit data within $gt \in [0,20]$.  (e) Lyapunov exponent as a function of $g$ and $\eta$ obtained from classical dynamics for the same initial state as (a)--(d).
} 
\label{figapp:nTWA}
\end{center}
\end{figure}

Fig.~\ref{figapp:nTWA}(a) shows the QFI $F(\hat S_{\rm opt})$ as a function of evolution time $gt$, obtained for a very large system of $N=10^8$. The initial state is taken to be identical as Fig.~2(b) of the main text: $(\theta_0,\phi_0,\alpha_0)=(1,0,0)$. For the larger system, we observe a much clearer exponential growth of $F(\hat S_{\rm opt})$ in the chaotic regime, up to the order of $N^2$ (over $8$ decades). The black dashed lines indicate the prediction $F\propto e^{2\lambda_L gt}$ for parameters $\tilde g=1.953$ and $\eta=1$ in the chaotic regime (green dashed dotted line). Note that for this comparison we use $\lambda_L$ obtained from a classical calculation instead of a fitted exponent. The observed agreement between the prediction based on $\lambda_L$ and the computed $F(\hat S_{\rm opt})$ is consistent with the results of Ref.~\cite{LewisSwan2019Scrambling}. We also show $F(\hat X_{\rm opt})$ for a boson displacement in Fig.~\ref{figapp:nTWA}(b) and observe similar exponential growth. Furthermore, we systematically investigate the growth of the QFI by fitting it to an exponential function $Ae^{\alpha_{S,X} gt}$. In Fig.~\ref{figapp:nTWA} (c) and (d), we plot the fitted parameter $\alpha_s$ and $\alpha_X$ for $F(\hat S_{\rm opt})$ and $F(\hat X_{\rm opt})$, respectively. We clearly observe a correspondence between the parameter regime of classical chaos [see Fig.~\ref{figapp:nTWA} (e) for Lyapunov exponent] and exponential growth of the QFI, with the latter indicated by $\alpha_{s,X} > 0$. Moreover, our results are quantitatively consistent with the prediction $\alpha_{s,X} \simeq 2\lambda_L$ of Ref.~\cite{LewisSwan2019Scrambling}.

\begin{figure}[bt]
\begin{center}
\includegraphics[width=1\columnwidth]{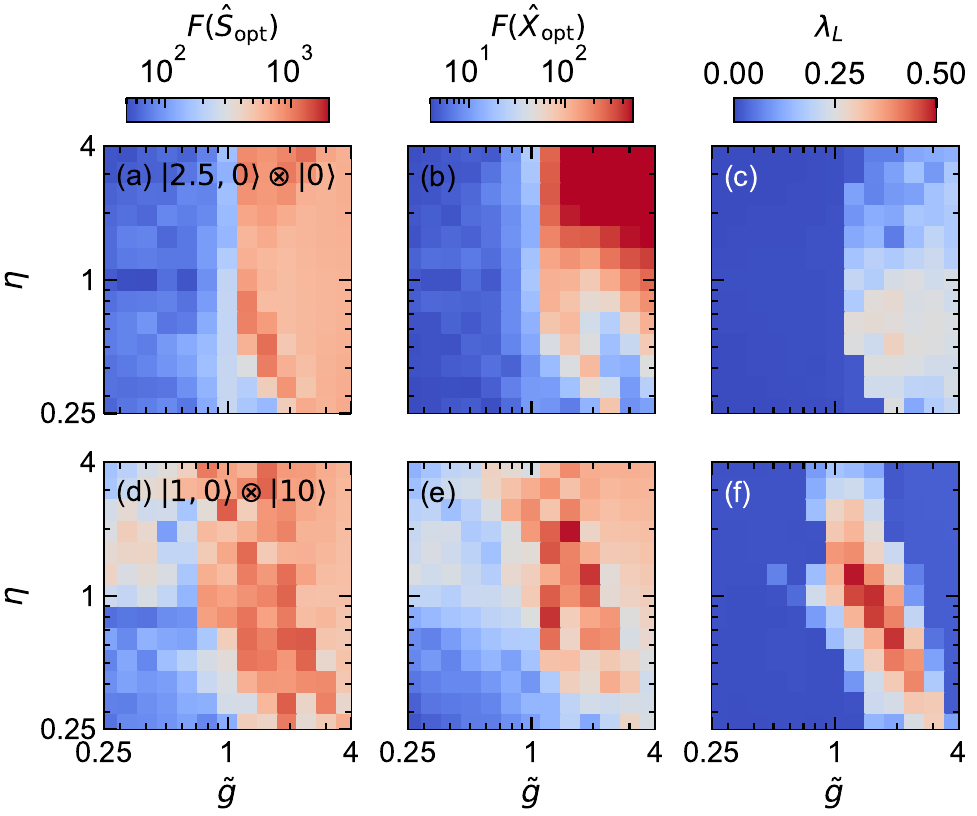}
\caption{{\it QFI for other initial states.} (a) and (b) The maximum QFI $F(\hat S_{\rm opt})$ and $F(\hat X_{\rm opt})$, respectively, for initial state with $(\theta_0,\phi_0,\alpha_0)=(2.5,0,0)$ at $gt=5$. Results are from exact calculations with $N=50$. (c) Corresponding value of Lyapunov exponent $\lambda_L$ obtained from classical dynamics with same initial state. Panels (d)---(f) are the same as (a)---(c) but for the initial state with $(\theta_0,\phi_0,\alpha_0)=(1,0,10)$ and at $gt=4$.
} 
\label{figapp:otherstate}
\end{center}
\end{figure}

\noindent{\bf Alternative initial conditions ---}
To demonstrate that our choice of ``generic" initial state $(\theta_0,\phi_0,\alpha_0)=(1,0,0)$ is not anomalous, we provide numerical data in Fig.~\ref{figapp:otherstate} for two other examples. The first is $(\theta_0,\phi_0,\alpha_0)=(2.5,0,0)$ which differs by changing the spin angle $\theta_0$. Panels (a)---(c) of Fig.~\ref{figapp:otherstate} show $F(\hat S_{\rm opt})$ and $F(\hat X_{\rm opt})$ at $gt=5$, and $\lambda_L$, respectively. Note that we sample data at a longer evolution time than the main text data, as the maximum value of $\lambda_L$ for this initial state is reduced and thus the scrambling time ($\sim \log(N)/\lambda_L$) is similarly larger. Comparing to Fig.~2 of the main text, we observe a similar level of correspondence between the regions of classical chaos and the behaviour of the QFI. We point out that for $F(\hat X_{\rm opt})$, there is a clear asymmetry in the QFI towards larger values for $\eta>1$. This is reconciled by noting that the bound of the QFI for bosonic displacements is intimately related to the occupancy of the bosons (the effective Heisenberg limit is approximately set by $(\Delta \Theta)^2 \sim 1/\langle \hat{a}^{\dagger}\hat{a}\rangle$), and we typically expect the population, and hence achievable QFI, to be larger when $\eta > 1$. We carry out a second calculation for the initial condition $(\theta_0,\phi_0,\alpha_0)=(1,0,10)$, which differs by featuring an initial non-zero boson occupancy. Results for the QFI (obtained at fixed $gt = 4$) and Lyapunov exponent are similarly shown in Figs.~\ref{figapp:otherstate}(d)---(f). Again, we observe a correspondence between the regions of classical chaos and the behaviour of the QFI, consistent with our prior conclusions.

\noindent{\bf Versatility for sensing ---}
In the main text, all results for the QFI are shown after optimization over the choice of rotation axis ($\hat n_{s}$) or displacement direction ($\varphi$). Operationally, we construct the QFI matrix $\mathbf{\Gamma}$ with elements $\Gamma_{ij}= 4\left(\langle\hat S_i\hat S_j+\hat S_j\hat S_i\rangle/2 -\langle\hat S_i\rangle\langle\hat S_j\rangle\right)$ ($i,j=x,y,z$) for the case of spin rotations and $\Gamma_{ij}=4\left(\langle\hat X_i\hat X_j+\hat X_j\hat X_i\rangle/2-\langle\hat X_i\rangle\langle\hat X_j\rangle\right)$ ($i,j=0,\pi/2$) for bosonic displacements. The maximal QFI $F(\hat S_{\rm opt})$ or $F(\hat X_{\rm opt})$ is given by the largest eigenvalue of ${\bm \Gamma}$ in either case \citeSI{Hyllus2010Fisher}. 

\begin{figure}[bt!]
\begin{center}
\includegraphics[width=1\columnwidth]{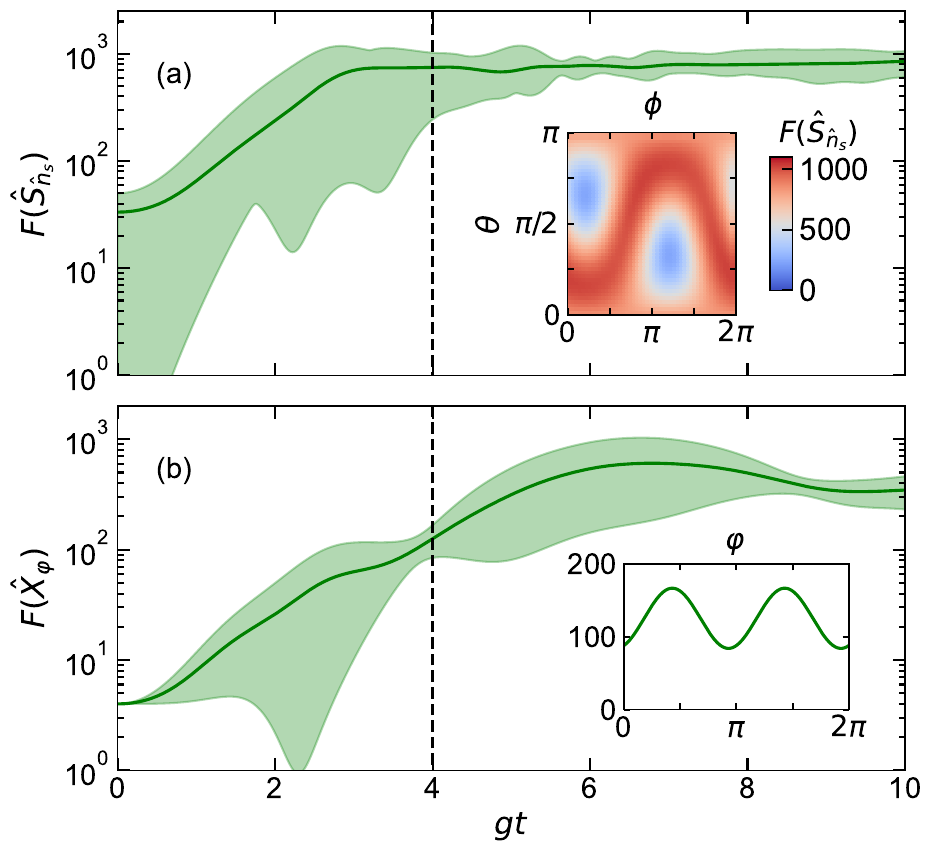}
\caption{{\it QFI for range of rotations and displacements.} (a) QFI $F(\hat S_{\hat n_s})$ for a spin rotation about a general axis $\hat n_s$ as a function of $gt$. The shaded region indicates the full range of QFI values obtained for arbitrary choice of $\hat n_s$, while the solid line shows the mean value obtained by averaging $F$ over all possible axes. (b) Analogous QFI $F(\hat X_{\varphi})$ for boson displacements along an arbitrary direction parametrized by $\varphi$ (see main text). In both panels (a) and (b), the results are for the initial state $(\theta_0,\phi_0,\alpha_0)=(1,0,0)$ in the regime of chaotic dynamics, $\tilde g=1.953$ and $\eta=1$, with $N=50$. Insets show the landscape of all possible QFI values at $gt=4$ (vertical dashed lines in the main panels).
} 
\label{figapp:QFIangle}
\end{center}
\end{figure}

An important benefit of using chaotic dynamics to generate entangled and correlated states for sensing is that the choice of generator, set by $\hat n_s$ and $\varphi$, does not need to be fine tuned to capture an anomalously large QFI or fast growth \cite{Shi2024chaos}. To demonstrate this, we show the dynamics of the QFI for all values of $\hat n_s$ or $\varphi$ in Fig.~\ref{figapp:QFIangle} using the initial state $(\theta_0,\phi_0,\alpha_0)=(1,0,0)$ with parameters $\tilde{g} = 1.953$ and $\eta = 1$ [identical to the chaotic example shown in Fig.~2(b) of the main text]. The green shaded areas indicate the range of the QFI for all possible generators, with the solid line indicating the average value. We observe that, identical to the behaviour of the maximal QFI shown in the main text, the average QFI also grows exponentially fast at short times and then saturates to $\sim N^2$ at long times. On the other hand, the shaded region becomes smaller and is minimal when the system becomes scrambled at long times. This is consistent with an analysis based on random matrix theory recently reported in Ref.~\cite{Shi2024chaos}. The insets in each panel of Fig.~\ref{figapp:QFIangle} similarly visualize the QFI at a fixed time of $gt = 4$ (approximately at the scrambling time for $N=50$) as a function of the parametrization angles $(\theta,\phi)$ and $\varphi$ of the spin rotation and boson displacement, respectively. We observe that the minimal QFI remains above the SQL in either case, which reinforces the robustness of the quantum-enhancement provided by chaotic dynamics.

\section{Time-reversal IBR}

In this section, we elaborate further on a number of technical aspects and details associated with the time-reversal IBR proposed in the main text. In particular, we further discuss the QFI attained when limited to measurements of only a single degree of freedom, possible extensions of our IBR protocol to account for limited readout, the impact of imperfect time-reversal and modelling of detection noise. 

\noindent{\bf QFI of subsystem ---}
We first introduce a rigorous definition for the spin (boson) QFI $\tilde F_{\mathrm{sp}}$ ($\tilde F_{\mathrm{b}}$) that we use in the main text. The spin (boson) QFI bounds the sensitivity that may be achieved when constrained to measurements of only spin (boson) observables, $(\Delta\Theta)^2 \geq \tilde F_{\mathrm{sp}}^{-1}$ ($\tilde F_{\mathrm{b}}^{-1}$). 

In general, spin-boson entanglement will lead to the reduced state of the spin (boson) degree of freedom being mixed. Thus the QFI of the reduced state with respect to a generator $\hat{G}$ is defined by,  
\begin{equation}\label{eq:QFI}
    \tilde F(\hat G)=4\lim_{\Theta\to0}\frac{1-{\cal F}(\hat G)}{\Theta^2}\, , 
\end{equation}
where 
${\cal F}=\bigg (\Tr\sqrt{\sqrt{\tilde \rho^{\Theta=0}}\tilde \rho^{\Theta}\sqrt{\tilde \rho^{\Theta=0}}}\bigg)^2$ is the Uhlmann fidelity 
and $\tilde \rho^{\Theta} = \mathrm{Tr}_{b(s)}\left[ \hat{U}_{\Theta}\vert\Psi_P\rangle \right]$ is the reduced density matrix of the spins (bosons) after unitary evolution $\hat{U}_{\Theta}$ parametrized by $\Theta$. For example, if the spin QFI is computed immediately after the spin rotation is applied then $\hat{U}_{\Theta} = e^{i\Theta\hat{S}_{\hat{n}_s}}$, while after time-reversal we instead use $\hat{U}_{\Theta} = e^{i\hat{H}_{D}t_{\mathrm{rev}}} e^{i\Theta\hat{S}_{\hat{n}_s}}$.

\begin{figure}[bt!]
\begin{center}
\includegraphics[width=1\columnwidth]{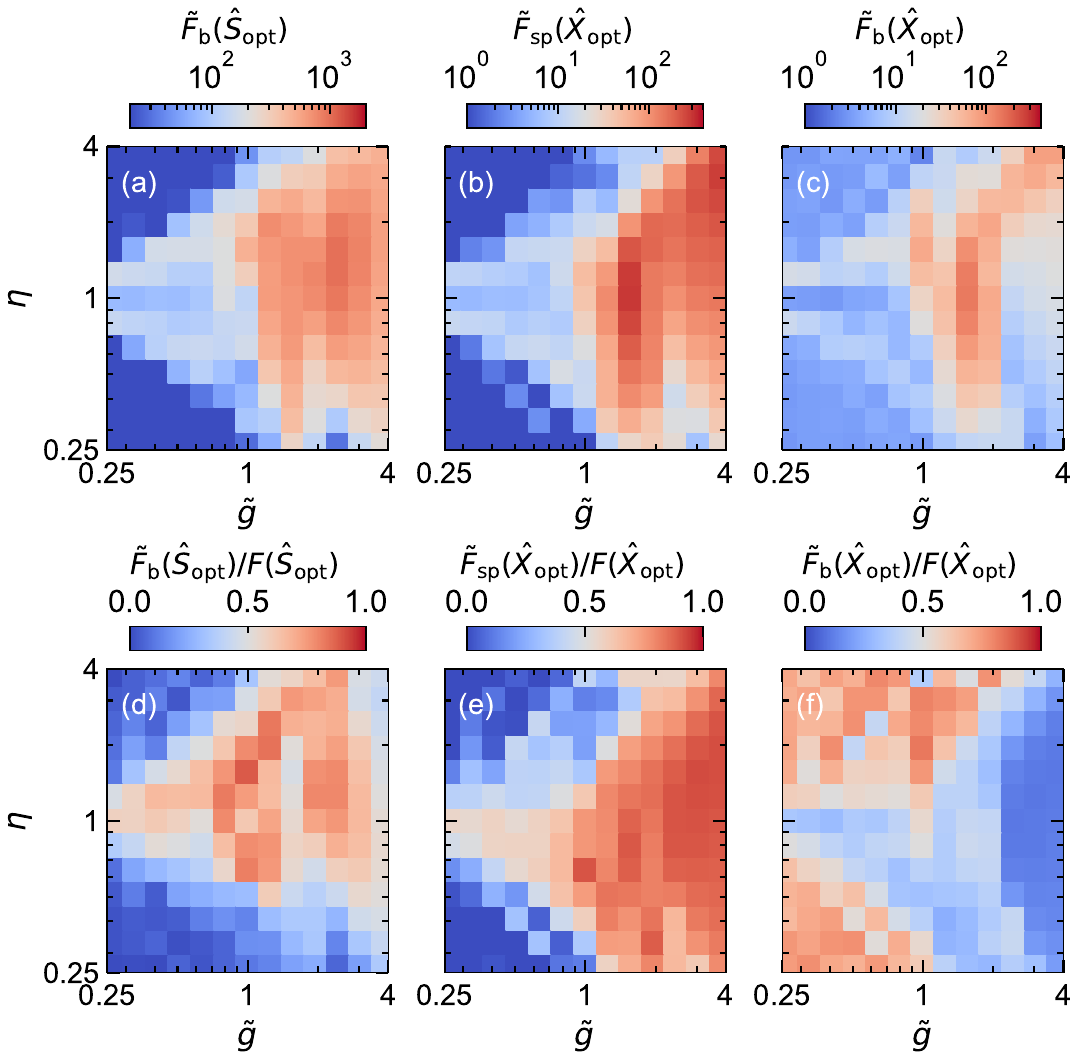}
\caption{{\it QFI constrained to individual degrees of freedom.} (a)--(c) Maximal spin and boson QFI, $\tilde{F}_{\mathrm{sp}}$ and $\tilde{F}_{\mathrm{b}}$, respectively, after time-reversal IBR for either: (a) spin rotation or (b), (c) boson displacement. All data is for an initial state $(\theta_0,\phi_0,\alpha_0)=(1,0,0)$ with $N = 50$ and obtained at fixed $gt = gt_{\rm rev} =4$. Axes indicate choice of parameters $\tilde{g}$ and $\eta$. (d)--(f) Equivalent results rescaled by the full QFI $F$ obtained from the complete spin-boson state.} 
\label{figapp:ReduceQFI}
\end{center}
\end{figure}

In Figs.~3(b) and (c) of the main text, we have shown the spin QFI $\tilde F_{\mathrm{sp}}$ and the ratio $\tilde F_{\mathrm{sp}}/F$ for an optimal spin rotation \footnote{We maximize the full QFI to obtain the optimal rotation axis.} $\hat G=\hat S_{\rm opt}$, respectively. We found that for balanced time reversal $t_{\rm rev}=t$ the spins and bosons become disentangled and the spin QFI can reach a large fraction of the QFI of the full system. Here, we provide analogous calculations of the boson QFI for a spin rotation , $\tilde F_b(\hat S_{\rm opt})$, in Figs.~\ref{figapp:ReduceQFI}(a) and (d). In the chaotic regime, we similarly observe that the boson QFI can become an appreciable fraction of the full QFI. For extreme values of $\eta$, the dominant QFI tends to follow naive expectations: When $\eta$ is small (spins dominate the dynamics) the spin QFI tends to be largest, while for large $\eta$ (bosons dominate the dynamics) the boson QFI is instead larger. However, for moderate values of $\eta$ the picture is more complicated and the precise details depend on the particular state and the generator of the perturbation.

We show analogous data for the spin and boson QFI with respect to an optimal boson perturbation $\hat{X}_{\mathrm{opt}}$ in panels (b), (e) and (c), (f) of Fig.~\ref{figapp:ReduceQFI}, respectively. While the fine details vary from the prior discussion, we again make the key observation that the time-reversal can enable quantum-enhanced sensitivity when limited to measurements of only a single degree of freedom (i.e., the spin and/or boson QFI can be a meaningful fraction of $F$).

\noindent{\bf State-transfer IBR for spin readout---}
In the main text and the prior paragraphs, we have demonstrated that a time-reversal IBR can enable quantum-enhanced sensing with measurements of only a single degree of freedom. Here, we briefly discuss an extension of our IBR protocol that can be useful in experimental systems where a specific degree of freedom may be inaccessible. A straightforward example are trapped ion arrays, where direct high-fidelity detection of the vibrational (boson) modes is typically not possible. On the other hand, the population of the qubit states of each ion is simply detected via fluorescence. In this scenario, a simple practical solution is to choose the operating  parameters $\tilde{g}$ and $\eta$ so that the spin QFI after time-reversal is large (see above). However, if this is not possible and the spin QFI becomes significantly reduced while the boson QFI is large, it is still possible to exploit the potential metrological advantage through a spin readout by appending an additional ``state transfer" IBR to the sequence of Fig.~1(a). 

Specifically, we propose to adopt an IBR developed in Refs.~\cite{Lewisswan2024Exploiting,Barberena2024Fast} that consists of evolution under a collective red sideband transition involving both the spin and boson degrees of freedom and described by the Hamiltonian,
\begin{equation}\label{eq:state-transfer-hamiltonian}
    \hat{H}_{\mathrm{rsb}} = \frac{g}{\sqrt{N}}\left(\hat{S}^+\hat{a} + \hat{S}^-\hat{a}^{\dagger}\right),
\end{equation}
where $\hat{S}^{\pm}$ are the raising and lowering operators of the collective spin. This Hamiltonian can either be realized directly or from the Dicke Hamiltonian \cite{Barberena2024Fast} by applying a sudden quench of the system parameters to the regime $\eta = 1$ and $\tilde{g} \ll 1$~\footnote{In this case, $\hat{H}_{\mathrm{rsb}}$ is realized in a rotating frame.}. The IBR is based on the insight that, in the limit of large $N$, Eq.~(\ref{eq:state-transfer-hamiltonian}) can be approximately recast as a beam-splitter Hamiltonian $\hat{H}_{\mathrm{sb}} \approx ig(\hat{b}^{\dagger}\hat{a} - \hat{b}\hat{a}^{\dagger})$ by using a Holstein-Primakoff transformation $\hat{S}^{+} \leftrightarrow \sqrt{N}\hat{b}^{\dagger}$ ($\hat{S}^{-} \leftrightarrow \sqrt{N}\hat{b}$) to treat the collective spin as an effective bosonic oscillator. Application of the beam-splitter Hamiltonian for a specific time can then be used to interchange the states of the coupled (effective) oscillators. Operationally, we propose to evolve the spin-boson system with $\hat{H}_{\mathrm{rsb}}$ for a time $gt = \pi/2$ at the conclusion of the time-reversal IBR. This ideally transfers the quadrature fluctuations of the boson mode onto the counterpart projection noise of the spin \cite{Lewisswan2024Exploiting}. The information about $\Theta$ encoded in the inaccessible boson degree of freedom prior to the state transfer IBR (signified by a large boson QFI) is then obtained via the measurement of spin observables after the state transfer IBR.

We note that in order to achieve perfect state transfer and thus metrological gain with this IBR there are some conditions specified within Refs.~\cite{Lewisswan2024Exploiting,Barberena2024Fast} that must be satisfied by the state at the conclusion of the preceding time-reversal sequence. Specifically, the boson state must be only weakly perturbed (either displaced or squeezed) from the vacuum state and the spin state must be strongly polarized along a specific direction (nominally along $\hat{z}$, although possibly defined with respect to a rotating frame). In the case where the boson QFI is large and the spin QFI is small, we expect that the spin state after time-reversal has returned very close to the initial condition $\ket{\theta_0,\phi_0}$, while the boson state is weakly perturbed (by an amount on the scale of the vacuum fluctuations, which encodes the information about $\Theta$) from the initial vacuum state. To satisfy the criteria for the state transfer IBR, it is thus simply necessary to first apply a global rotation (about a known axis and angle determined from $\theta_0$ and $\phi_0$) of the qubits to orient them along the appropriate direction, i.e., $\hat{z}$, and then continue with the state transfer IBR. 

\noindent{\bf Optimal spin measurement ---}
In the main text, we state that the projection $\hat P_s=\ket{\theta_0,\phi_0}\bra{\theta_0,\phi_0}$ is a sufficient measurement to saturate the bound established by the spin QFI, $(\Delta\theta)^2 = \mathrm{Var}(\hat{P}_s) /\vert \partial_{\Theta} \hat{P}_s \vert^2 = \tilde{F}_{\mathrm{sp}}^{-1}$, with balanced time-reversal. The proof of this statement follows analogously to a similar result in Ref.~\cite{Macri2016Loschmidt}, wherein it is shown that $\hat{P} = \ket{\Psi_0}\bra{\Psi_0}$ (i.e., projection into the full spin-boson initial state) saturates the QCRB after time-reversal. As the final state in the absence of a perturbation is an uncorrelated product state (i.e., $\ket{\Psi_F^{\Theta=0}} = \ket{\Psi_0}$), the Uhlmann fidelity in Eq.~(\ref{eq:QFI}) can be written as ${\cal F}= \langle\theta_0,\phi_0|\tilde \rho^{\Theta}|\theta_0,\phi_0\rangle = \langle \hat{P}_s \rangle$ where $\tilde \rho^{\Theta}=\Tr_b[|\Psi^{\Theta}_F\rangle\langle\Psi^{\Theta}_F|]$. Then, using Eq.~\eqref{eq:QFI}, we have $\tilde F_{\rm sp}=4\lim_{\Theta=0}\frac{1-\langle\hat P_s\rangle}{\Theta^2}$ or equivalently $\langle\hat P_s\rangle = 1 - (\tilde F_{\rm sp}/4)\Theta^2 + \mathcal{O}(\Theta^3)$. The inverse sensitivity for a measurement of $\hat P_s$ then becomes,
\begin{align}
    \frac{1}{(\Delta\Theta)^2}\bigg|_{\Theta\to0}=&\frac{|\partial\langle\hat P_s\rangle/\partial\Theta|^2}{{\rm Var}(\hat P_s)}\bigg|_{\Theta\to0}\nonumber\\=&\frac{|\partial\langle\hat P_s\rangle/\partial\Theta|^2}{\langle\hat P_s\rangle(1-\langle\hat P_s\rangle)}\bigg|_{\Theta\to0}\nonumber\\=&\frac{|(-\Theta/2)\tilde F_{\rm sp}|^2}{(\Theta^2/4)\tilde F_{\rm sp}}=\tilde F_{\rm sp}\,,
\end{align}
where we have assumed we are working in the limit of $\Theta\to0$ and have used the relation ${\rm Var}(\hat P_s)=\langle\hat P_s\rangle(1-\langle\hat P_s\rangle)$.

\noindent{\bf Imperfect time reversal ---}
In Fig.~3(e) of the main text we showed results for the dependence of the inverse sensitivity $1/(\Delta \Theta)^2$ on the reversal time $t_{\rm rev}$, for a number of different measurement observables. The data shown in that figure was obtained at a fixed $\Theta=1/N$, but here we briefly discuss the dependence of the attainable sensitivity with a general (imbalanced) time-reversal IBR on the operating rotation angle $\Theta$. 

\begin{figure}[bt!]
\begin{center}
\includegraphics[width=1\columnwidth]{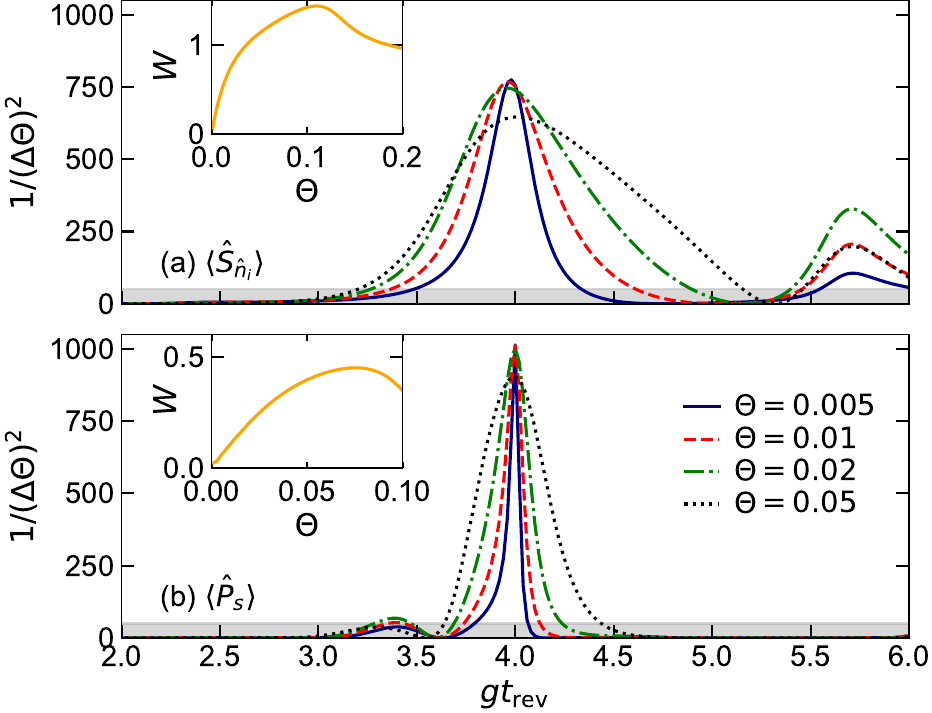}
\caption{{\it Sensing performance with imperfect time reversal.} Inverse sensitivity $1/(\Delta \Theta)^2$ with a measurement of (a) $\hat S_{\hat n_i}$ or (b) $\hat{P}_s$, as a function of reversal time $t_{\rm rev}$. We distinguish the sensitivity obtained with different working angles $\Theta$ (see legend). All data uses an initial state $(\theta_0,\phi_0,\alpha_0)=(1,0,0)$ with $N=50$, Hamiltonian parameters $\tilde g=1.953$ and $\eta=1$, and initial entangling time $gt=4$.Insets: Full-width-at-half-maximum $W$ of the peak centered about $gt_{\rm rev}\approx4$ as a function of $\Theta$.} 
\label{figapp:sensitivity}
\end{center}
\end{figure}

Figure \ref{figapp:sensitivity} shows the inverse sensitivity (for an optimal spin rotation, $\hat{G} = \hat{S}_{\mathrm{opt}}$) as a function of $t_{\mathrm{rev}}$ for (a) $\hat M=\hat S_{\hat n_i}$ and (b) $\hat M=\hat P_s$. The Hamiltonian parameters ($\tilde{g} = 1.953$ and $\eta = 1$) and evolution time $gt = 4$ is chosen to match Fig.~3(e) of the main text. We compare results for working angles $\Theta=0.005$ (blue solid line), $\Theta=0.01$ (red dashed line), $\Theta=0.02=1/N$ [green dashed dotted line, same as Fig.~3(e)], and $\Theta=0.05$ (black dotted line). We observe two clear trends in the results: i) As $\Theta$ deviates further from zero, the optimal sensitivity slightly reduces in both cases, and is also located away from the balanced time-reversal IBR ($t_{\rm rev}=t$) for $\hat{M}= \hat{S}_{\hat{n}_i}$. ii) At the same time, the marginal reduction in sensitivity is accompanied by a much more pronounced broadening of quantum-enhanced performance (performance worse than the SQL is indicated by the grey shaded region) as $\Theta$ increases in magnitude. To emphasize the latter, in the inset of each panel we show the full-width-at-half-maximum $W$ of the central feature (defined with respect to the maximum inverse sensitivity) as a function of $\Theta$. The fact that $W$ appears to vanish for small $\Theta$ can be of practical relevance, and suggests that a preferable strategy is to work with small but finite $\Theta \neq 0$ and marginally sacrifice absolute sensitivity performance for a more robust tolerance to timing issues. We note that such a strategy is in fact already desirable due to the inherent difficulty of working at a point dominated by technical noise (the variance and signal both vanish as $\Theta \to 0$ for $\hat{M} = \hat{P}_s$ and $\hat{S}_{\hat{n}_i}$). It is for these reasons that we use results obtained with $\Theta = 1/N$ in both Fig.~3(e) and Fig.~4 of the main text.

\noindent{\bf Modeling of technical imperfections ---}
In the main text we consider the robustness of our proposed sensing scheme to technical imperfections such as detection noise, Hamiltonian parameter fluctuations and imperfect state preparation. Here, we provide further detail of the specific models that we adopt in each case.

\noindent{\it Parameter fluctuations:} Apart from timing issues, imperfect time-reversal can occur due to uncertain calibration or shot-to-shot fluctuations in Hamiltonian parameters. In our case, we focus on shot-to-shot variation of the boson frequency $\delta$ as this has shown to be relevant in large $2$D ion crystals used to simulate the Dicke model \cite{Gilmore2021Quantum,Safavi2018Verification}. In the main text, we show results using a model where the time-reversal IBR is implemented by abruptly jumping the parameters of the Dicke Hamiltonian so that $g \to g_{\mathrm{rev}} = - g$, $\Omega \to \Omega_{\mathrm{rev}} = - \Omega$ and $\delta \to \delta_{\mathrm{rev}}$. While the sign of the former pair of parameters is perfectly reversed, we assume that the boson frequency varies between experimental shots according to the Gaussian distribution,\begin{equation}
    {\cal P}(\delta_{\rm rev})=\frac{1}{\sqrt{2\pi\sigma_{\delta}^2}}e^{-\frac{[\delta_{\rm rev}-(-\delta)]^2}{2\sigma_{\delta}^2}}\, .
\end{equation}
Here, $\delta_{\mathrm{rev}}$ has the correct mean value $-\delta$ but features rms fluctuations $\sigma_{\delta}$. Results in Fig.~4(a) of the main text are obtained by simulating the full time-reversal sequence with $\delta_{\mathrm{rev}}$ sampled from the above distribution and constructing relevant expectation values. The perturbation $\hat G=\hat S_{\rm opt}$ is the optimized spin rotation chosen to maximize the QFI before subsequent time-reversal. 

\noindent{\it Imperfect state preparation:} In Fig.~4(b) of the main text we model imperfect state preparation by assuming that the boson mode is not properly cooled to the vacuum state initially. This can be efficiently treated by simulating the dynamics of an ensemble of Fock state initial conditions, $\ket{\Psi_0^{n}} = \ket{\theta_0,\phi_0} \otimes \ket{n}$ where $\ket{n}$ an eigenstate of the number operator $\hat{a}^{\dagger}\hat{a}\ket{n} = n\ket{n}$. As an initial thermal state of the bosons does not feature coherences between different Fock components, we may construct relevant observables at the conclusion of our sensing protocol as $\langle \hat{O}\rangle = \sum_{n=0}^{n_{\mathrm{cut}}} {\cal P}_n \langle \hat{O} \rangle_n$ where $\langle \hat{O} \rangle_n$ is the expectation value obtained from an initial state $\ket{\Psi_0^{n}}$ and 
\begin{equation}
    {\cal P}_n=\frac{(\bar {\cal N})^n}{(\bar {\cal N}+1)^{n+1}} ,
\end{equation}
with $\bar {\cal N}$ is the average occupancy of the initial thermal boson state. We choose the cut-off $n_{\mathrm{cut}}$ for a given choice of $\bar {\cal N}$ by checking the convergence of relevant (thermal) expectation values. 

\noindent{\it Detection noise:} In Fig.~4(c) of the main text we show the effects of imperfection detection of spin observables. Specifically, we show the impact of detection noise on a measurement of the spin projection $\hat{M} = \hat{S}_{\hat{n}_i}$, as well as a more sophisticated strategy using the classical Fisher information (CFI). 

For a measurement of $\hat S_{\hat n_i}$, we can model detection noise by adding a Gaussian random variable $\xi_M$, with zero mean and variance $\sigma^2_M$, to each measurement outcome (i.e., shot of an experiment) \cite{Guan2021Tailored}. This leaves the mean spin projection unchanged, $\langle \hat S_{\hat n_i}\rangle_{\sigma_M}=\langle \hat S_{\hat n_i}\rangle_{\sigma_M=0}$, but the variance transforms as ${\rm Var}(\hat S_{\hat n_i})_{\sigma_M}={\rm Var}(\hat S_{\hat n_i})_{\sigma_M=0}+\sigma_M^2$. These expectation values are then used to evaluate the sensitivity, $(\Delta\Theta)^2 = \mathrm{Var}(\hat S_{\hat n_i})_{\sigma_M}/\vert \partial_{\Theta}\langle \hat S_{\hat n_i} \rangle_{\sigma_M}\vert^2$.

The CFI $F^{\hat{M}}_C(\theta)$ provides a bound on the sensitivity given knowledge of the full counting statistics for a specific measurement $\hat{M}$, $(\Delta\Theta)^2 \geq (F^{\hat{M}}_C)^{-1}$. For a measurement of a spin projection along the arbitrary axis $\hat{n}$, $\hat{M} = \hat{S}_{\hat{n}}$, the CFI is defined as, 
\begin{equation}\label{eq:CFI}
    F^{\hat{S}_m}_C(\theta)=\sum_{m=-N/2}^{N/2} p_m(\Theta)\bigg[\frac{\partial p_{m}(\Theta)}{\partial\Theta}\bigg]^2\,,
\end{equation}
where $p_m=|\langle m_{\hat n}|\Psi_F\rangle|^2$ and $|m_{\hat n}\rangle$ is an eigenstate of $\hat{S}_{\hat{n}}$ with eigenvalue $m_{\hat{n}}$. Detection noise is incorporated by convolving $p_m$ with a Gaussian function of rms width $\sigma_M$ and mean zero to obtain,
\begin{equation}
    \tilde p_{m}=\frac{1}{\sqrt{2\pi\sigma_M^2}}\sum_{m'}p_{m}e^{-\frac{(m-m')^2}{2\sigma_M^2}}\, .
\end{equation}
Substitution of the convolved distribution into Eq.~\eqref{eq:CFI} yields an updated CFI that can be used to bound the achievable sensitivity. We note that the convolution reduces to the above treatment of Gaussian detection noise for the lowest moments of the spin projection when one identifies $\hat{n} \to \hat{n}_i$. 

In Fig.~4(c) of the main text, we plot the CFI obtained via this procedure as an indicator of the best attainable (inverse) sensitivity as a function of detection noise $\sigma_M$. We note that for every value of $\sigma_M$ we optimize the measurement axis $\hat n$, with the abrupt kinks observed in the solid lines corresponding to sudden changes of the optimal $\hat n$. Typically, we find that $\hat{n} = \hat{n}_i$ for $\sigma_M \lesssim 0.5$, i.e., where detection at the level of single spins can be achieved, while $\hat{n}$ is perpendicular to $\hat{n}_i$ for $\sigma_M \gtrsim 1$.

\bibliographystyleSI{apsrev4-1}
\bibliographySI{Reference}

\end{document}